\documentclass[12pt]{article}
\usepackage{amssymb}

\renewcommand{\t}{\theta}
\def\be{\begin{equation}}
\def\ee{\end{equation}}
\def\ba{\begin{eqnarray}}
\def\ea{\end{eqnarray}}
\newcommand{\nn}{\nonumber}
\newcommand{\no}{\nonumber \\}
\def\starcom {\stackrel{\star}{,}}
\renewcommand{\d}{\delta}
\def\D {\Delta}
\def\del{\partial}
\def\A{{\cal A}}
\def\ha{\frac{1}{2}}
\begin{document}
\begin{titlepage}
April 2003     
\vskip -0.55cm 
\hfill hep-th/0304030

\begin{center}

\vskip .15in

\renewcommand{\thefootnote}{\fnsymbol{footnote}}
{\large \bf The Seiberg-Witten Map for a Time-dependent Background}
\vskip .25in
B. L. Cerchiai\footnote{email address: bianca@physics.unc.edu}, 
\vskip .25in

{\em    Department of Physics  and Astronomy\\
        University of North Carolina\\
        296 Phillips Hall, CB\#3255\\
        Chapel Hill, NC 27599-3255}
\end{center}
\vskip .25in

\begin{abstract}

\end{abstract}

In this paper the Seiberg-Witten map for a time-dependent background
related to a null-brane orbifold is studied. The commutation
relations of the coordinates are linear, i.e. it is an example of the
Lie algebra type. The equivalence map between the Kontsevich star product for
this background and the Weyl-Moyal star product for a background with
constant noncommutativity parameter is also studied.
\end{titlepage}

\newpage
\section{Introduction}

Recently there has been much interest in time-dependent backgrounds
\cite{FS, LMS1, LMS2, CCK, HoPo, HaSe, DRRS, AlPa, DoNa, CLO, BaHu, LNR}.
One of the examples which has been studied is a null-brane orbifold
\cite{FS, LMS1, LMS2} of the four-dimensional
Minkowski space. In \cite{HaSe} Sethi and Hashimoto
compute the noncommutativity parameter $\t^{ij}$ for such a background, by
following the procedure introduced in \cite{CDS, DoHu} to compute the
commutation relations of the space-time coordinates. The string derivation
of this is in \cite{Sch, SW}.
It turns out that in this example $\t^{ij}$ has a linear dependence on
the space-time coordinates, i.e. it defines an algebra of the Lie algebra type.
It is an interesting example, since it is an algebra with particularly
simple properties, for which one of the coordinates is central and the
higher commutators of any element vanish. This makes it possible to perform
many computations explicitly.

This paper is divided as follows. In section 2 the algebra \cite{HaSe} 
is recalled and the corresponding Kontsevich star product \cite{Kon},
which generalizes the Weyl-Moyal star product when $\t^{ij}$ is not constant, 
is calculated by the Weyl quantization procedure \cite{Moy, Gro, MSSW}.
In section 3 an equivalence map \cite{Kon} in the sense of deformation 
quantization of this star product with the Weyl-Moyal star product for a 
certain algebra with constant $\t^{ij}$ is constructed. This is possible
because for the null-brane orbifold there is a coordinate transformation
\cite{LMS1, LMS2} relating the two descriptions.
Finally, in the last section the Seiberg-Witten map \cite{SW}, which expresses
a gauge theory defined on a noncommutative space-time in terms of a
corresponding commutative gauge theory, is computed for this background to
the lowest non-trivial orders in $\t^{ij}$. It is verified that 
while there are no corrections due to the time-dependence of $\t^{ij}$
for the gauge parameter $\Lambda$ to the first order in $\t^{ij}$, there are to
the second order.
For the gauge field $a_i$ there are time-dependent corrections already 
to the first non-trivial order in $\t^{ij}$. In order to obtain these 
results the cohomological method discussed in \cite{BCPVZ, BCZ} is used.

\setcounter{footnote}{0}
\section{The algebra and the star product}

In this paper the Seiberg-Witten map for the four-dimensional
noncommutative time-dependent background obtained in \cite{HaSe},
\cite{DRRS} by T-duality of a null-brane orbifold \cite{FS, LMS1, LMS2}
is studied. 

The algebra $\A$ formed by the coordinates is generated by
$\{x^+,x^-,x,z\}$ with relations
\be
[x^i,x^j]=i \: \theta^{ij}
\label{defx}
\ee
where $x^i \in \A$, for $i=1,\ldots, 4$, i.e. $x^1=x^+, \, 
x^2=x^-, \, x^3=x, \, x^4=z$,
\be
\t^{xz}=-\t^{zx}=\widetilde R x^+, \qquad \t^{x^-z}=-\t^{zx^-}=\widetilde R x
\label{deft}
\ee
and all the other components of $\t^{ij}$ vanish. Here $\widetilde R$ is 
constant and the orbifold identifications are
\ba
x^+ &\sim &x^+; \no
x\:\:\: &\sim &x+2 \pi x^+; \label{ident}\\
x^- &\sim & x^-+2 \pi x+\ha (2 \pi)^2 x^+; \no
z\:\:\:&\sim & z+\frac{2\pi}{\widetilde R}. \nn
\ea 
The algebra defined by (\ref{defx}),(\ref{deft}) is of the Lie algebra
type, since the commutation relations of the coordinates are linear and the
Jacobi identity is satisfied.
Moreover, it is nilpotent (thus solvable), in the sense that the third
commutator of any four elements of $\A$ vanishes
\be
[x^i,[x^j,[x^k,x^l]]]=0 \quad \forall x^i \in \A.
\label{nil}
\ee
A further observation is that $x^+$ is in the center of $\A$.
For these reasons it is a particular interesting example, because these
properties ensure that many computations can be actually carried out
explicitly.

As a first step, the explicit formula for the Kontsevich star product
\cite{Kon} corresponding to this algebra is calculated, by using the Weyl
quantization procedure \cite{Moy, Gro, MSSW}. 
The result, which will be derived below in (\ref{fourier})-(\ref{BCH}), is
\ba
f \star g&=&f \: exp\left( \frac{i}{2} \widetilde R x^+(\stackrel {\leftarrow}
\del_x \: \stackrel {\rightarrow} \del_z-\stackrel {\leftarrow} \del_z \:
\stackrel {\rightarrow} \del_x)
+\frac{i}{2} \widetilde R x (\stackrel {\leftarrow} \del_{x^-}
\stackrel {\rightarrow} \del_z-
\stackrel {\leftarrow} \del_z \: \stackrel {\rightarrow} \del_{x^-}) \right. 
\label{prod} \\
&&+\left. \frac{1}{12} \widetilde R^2 x^+(\stackrel {\leftarrow} \del_{x^-} 
\stackrel {\leftarrow} \del_z \: \stackrel {\rightarrow} \del_z
-\stackrel {\leftarrow} \del_z \: \stackrel {\leftarrow} \del_z \:
\stackrel {\rightarrow} \del_{x^-}-\stackrel {\leftarrow} \del_{x^-} 
\stackrel {\rightarrow} \del_z \: \stackrel {\rightarrow} \del_z
+\stackrel {\leftarrow} \del_z \: \stackrel {\rightarrow} \del_{x^-} 
\stackrel {\rightarrow} \del_z)\right) \: g. \nn
\ea
In order to compare this result with Kontsevich's formula explicitly, it can
be seen that to the second order in $\t^{ij}$ (\ref{prod}) reduces to
\ba
f \star g&=&f \: g+\frac{i}{2} \widetilde R x^+ \left( \del_x f \: \del_z g 
-\del_z f \: \del_x g \right) +\frac{i}{2} \widetilde R x \left(\del_{x^-} f \:
\del_z g -\del_z f \: \del_{x^-}g  \right) \no
&&-\frac{1}{8} \widetilde R^2 (x^+)^2 \left( \del^2_x f \: \del^2_z g +\del^2_z f 
\: \del^2_x g -2 \del_x \del_z f \: \del_x \del_z g \right) \no
&&-\frac{1}{8} \widetilde R^2 x^2 \left( \del^2_{x^-} f \: \del^2_z g 
+\del^2_z f \: \del^2_{x^-} g -2 \del_{x^-} \del_z f \: \del_{x^-} \del_z g
\right)\\
&&+\frac{1}{4} \widetilde R^2 x^+ x \left(-\del_{x^-} \del_x f \: \del^2_z g
+\del_x \del_z f \: \del_{x^-} \del_z g \right.+ \del_{x^-} \del_z f \: 
\del_x \del_z g \no
&&\left. -\del^2_z f \:\del_x\del_{x^-}g\right)+\frac{1}{12} 
\widetilde R^2 x^+\left(\del_{x^-} \del_z f \: \del_z g \right. \no
&& \left.-\del^2_z f \: \del_{x^-} g -\del_{x^-} f \: \del^2_z g
+\del_z f\: \del_{x^-} \del_z g)\right) +...       \nn
\ea
which coincides with Kontsevich' s expression \cite{Kon} to this order.
\ba
f \star g&=&f \: g+\frac{i}{2} \t^{ij} \del_i f \del_j g \\
&&-\frac{1}{8} \t^{ij} \t^{kl} \del_i \del_k f \del_j \del_l g
-\frac{1}{12} \t^{ij} \del_j \t^{kl} \left(\del_i \del_k f \del_l g -
\del_k f \del_i \del_l g\right)+... \nn
\label{kont}
\ea
The star product (\ref{prod}) is associative and the $\theta^{ij}$
appearing in (\ref{deft}) satisfies the Jacobi identity \cite{HaSe}
\be
\t^{ij} \del_j \t^{kl}+\t^{kj} \del_j \t^{li}+\t^{lj} \del_j \t^{ik}=0.
\label{jacobi}
\ee

In this particular example the star product (\ref{prod}) can be computed by
simply applying the Weyl quantization procedure \cite{Moy, Gro}, following the
method of \cite{MSSW}. Starting from the Fourier transform of a function
$f(x^i)$, with $x^i$ commutative variables
\be
\tilde f(k)=\int dx e^{-i k_j x^j} f(x)
\label{fourier}
\ee
the Weyl operator associated to $f(x^i)$ is defined as
\be
W(f)=\int dk e^{i k_i \hat x^i} \tilde f(k)
\ee
where the commutative variable $x^i$ is replaced with $\hat x^i \in \A$.
In this way a particular ordering of the elements $\hat x^i \in \A$ is
picked, i.e. the most symmetric one. Moreover, if the product of two
such operators $W(f)W(g)$ is considered
\be
W(f)W(g)=\int dk \: dp e^{i k_i \hat x^i} e^{i p_j \hat x^j} \tilde f(k)
\: \tilde g(p),
\ee
then the star product $f \star g$ can be constructed as the 
function corresponding to $W(f)W(g)$, i.e.
\be
f \star g=\int dk \: dp \:  e^{i (k_j+p_j+g_j(k,p)) x^j} \tilde f(k) \: 
\tilde g(p)~,
\label{weyl}
\ee
where the expression of $g_j(k,p)$ is obtained through the
Baker-Campbell-Hausdorff formula for the product of two exponentials
\be
e^A e^B=e^{(A+B+\ha [A,B]+\frac{1}{12} [A-B,[A,B]]+...)}
\label{BCH}
\ee
applied to $A=k_i \hat x^i,B=p_j \hat x^j$.

For the algebra $\A$ defined in (\ref{defx}) and (\ref{deft}) the triple
commutator of any four elements vanishes (\ref{nil}).
Therefore in the formula (\ref{BCH}) the contribution from the ellipses
vanishes and the result (\ref{prod}) for the star product is obtained.

\section{The coordinate transformation and equivalence map}

For the example (\ref{defx}), (\ref{deft}) an equivalence map to a background
with constant $\t^{ij}$ is now constructed.
According to \cite{LMS1},\cite{LMS2},\cite{HaSe} the same background
described by (\ref{defx}),(\ref{deft}) can be described also
in terms of another algebra generated by the elements $\{y^+,y^-,\tilde y,z\}$
with commutation relations
\be
[y^i,y^j]=i \: \tilde \theta^{ij}
\ee
where $y^1=y^+, \, y^2=y^-, \, y^3=\tilde y, \, x^4=z$,
\be
\tilde \theta^{\tilde y z}=-\tilde \theta^{z \tilde y}=\widetilde R
\label{tildet}
\ee
for $\widetilde R$ constant, and the other components of 
$\tilde \t^{ij}$ vanish. 

The map $\sigma$ relating the two algebras is
\be
\begin{array}{lcl}
x^+&=&y^+; \\
x &=&y^+\left( \tilde y+\widetilde R z \right); \\
x^-&=&y^-+\ha y^+ \left(\tilde y+\widetilde R z\right)^2.
\end{array}
\label{transf}
\ee
The orbifold identification (\ref{ident}) in the new variables becomes
\be
\begin{array}{lcl}
y^+ & \sim & y^+; \\
\tilde y& \sim & \tilde y+2\pi; \\
y^- &\sim& y^-; \\
z &\sim& z+{\displaystyle \frac{2 \pi}{\widetilde R}}.
\end{array}
\ee
It can be noticed that the coordinate transformation (\ref{transf})
is not linear and that it is singular for $x^+=y^+=0$.

According to Kontsevich's formality theorem \cite{Kon}, where $\sigma$ is
well-defined, the star products $\star$ and $\tilde \star$ corresponding
to $\t^{ij}$ in (\ref{deft}) and $\tilde \t^{ij}$ in (\ref{tildet}) respectively
are equivalent up to the coordinate transformation $\sigma$.

This means that if the coordinate transformation (\ref{transf}) is applied to
the Weyl-Moyal product
\be
f \tilde \star g= f e^{\frac{i}{2} \widetilde R
(\stackrel{\leftarrow}{\del_{\tilde y}}
\stackrel{\rightarrow}{\partial_z}-\stackrel{\leftarrow}{\del_z}
\stackrel{\rightarrow}{\del_{\tilde y}})} g
\ee
associated to $\tilde \t^{ij}$, then the new star product
\be
f \star' g=f \star g-\frac{1}{24}\widetilde R^2 x^+ \left(\del_{x^-}f \del^2_z
g+ \del^2_z f \del_{x^-} g+2 \del_z \del_{x^-} f \del_z g+2 \del_z f 
\del_z \del_{x^-} g\right)+...
\label{pstar}
\ee
has to be equivalent to $\star$ defined in (\ref{prod}).

Notice that to obtain (\ref{pstar}) the following relation
\be
\del_{\tilde y}=x^+\del_x+x\del_{x^-}
\ee
has been used, which follows from (\ref{transf}).

Two star products are equivalent if there exists an equivalence map 
$R$, i.e. a differential operator, such that
\be
f \star' g=R^{-1} \left( R(f) \star R(g)\right).
\label{equiv}
\ee

$R$ and the star products are expanded in powers of $\theta$
\be
R(f)=f+R^{(1)}(f)+R^{(2)}(f)+...
\ee
and
\be
\begin{array}{lcl}
f \star g&=& fg+B^{(1)}(f,g)+B^{(2)}(f,g)+... \\
f \star' g&=& fg+{B'}^{(1)}(f,g)+{B'}^{(2)}(f,g)+...
\end{array}
\ee
Here $R^{(n)}$, $B^{(n)}$, ${B'}^{(n)}$ denote the contribution of order
$n$ in $\t^{ij}$.
Then (\ref{equiv}) becomes
\be
\begin{array}{lcl}
{B'}^{(1)}(f,g)&=&B^{(1)}(f,g)+R^{(1)}(f) \, g+f \, R^{(1)}(g)
-R^{(1)}(f \, g)\, ,\\ \\
{B'}^{(2)}(f,g)&=&B^{(2)}(f,g)+R^{(2)}(f) \, g+f \, R^{(2)}(g)-R^{(2)}(f \, g) 
\\\\
&&-R^{(1)}(B^{(1)}(f,g))+R^{(1)}(R^{(1)}(f \, g))+R^{(1)}(f) \, R^{(1)}(g)\\ \\
&&+B^{(1)}(R^{(1)}(f),g)+B^{(1)}(f,R^{(1)}(g)) \, .
\end{array}
\ee

In this case the equivalence map to the fourth order in $\t^{ij}$ is 
found to be
\ba
R^{(1)}(f)=0, && R^{(2)}(f)=-\frac{1}{24} \widetilde R^2 x^+ \del_{x^-}
\del^2_z f=\frac{1}{24} \t^{zx} \del_x \t^{x^-z} \del_{x^-}
\del^2_z f, \no
R^{(3)}(f)=0, && R^{(4)}(f)=\frac{1}{1152} \widetilde R^4 (x^+)^2 
\del^2_{x^-} \del^4_z f=\ha (R^{(2)})^2(f),
\ea
which suggests that the equivalence map is actually generated by the flow 
of $R^{(2)}$, i.e.
\be
R \: f=e^{-\frac{1}{24} \widetilde R^2 x^+ \del_{x^-} \del^2_z} \: f\:.
\ee
It can be seen that it is singular for $x^+=y^+=0$.

The equivalence of the star product (\ref{prod}) to the Weyl-Moyal product
$\tilde \t^{ij}$ is what guarantees that it is associative and hence that
the Jacobi identity (\ref{jacobi}) is satisfied. 

\section{The Seiberg-Witten map}

The Seiberg-Witten (SW) map \cite{SW} relating the commutative
and noncommutative gauge theories for the star product (\ref{prod})
is now derived. In order to achieve this, the covariant coordinates
\be
X^i=x^i+A^i(x^j), \qquad i=1,\ldots,d
\label{shift}
\ee
are introduced, according to \cite{Sei} and \cite{JuSch, JSW, MSSW}.
Here $d$ is the space-time dimension, in this case $d=4$.
The name covariant coordinates is justified by the observation that they
are required to transform like
\be
\delta X^i=i[\Lambda \starcom X^i] \equiv i (\Lambda \star X^i-X^i \star \Lambda)
\label{cov}
\ee
under an (infinitesimal) noncommutative gauge transformation $\delta$
with gauge parameter $\Lambda$.
The gauge potential $A^i$ in (\ref{shift}) is required to transform like
\be
\delta A^i=i [ \Lambda \starcom x^i]+i [ \Lambda \starcom A^i].
\label{swa}
\ee
It is a non-trivial result \cite{MSSW} that for the case of $\t^{ij}$ in
(\ref{deft}), which is linear, it is consistent
to identify $i [ \Lambda \starcom x^i]=\t^{ij} \del_j \Lambda$
in (\ref{swa}), because it is possible to write
$[x^i \starcom f]=i \, \t^{ij} \del_j f$ for any $f(x^i)$, where
the Jacobi identity is used to verify the Leibniz rule, and the index of the
derivative is raised with $\t^{ij}$.

The eqns. (\ref{cov}) and (\ref{swa}) guarantee that for a scalar field 
$\Psi(x^i)$ transforming as $\delta \Psi=0$ the following is true
\be
\delta (X^i \star \Psi)=i \Lambda \star (X^i \star \Psi).
\label{psi}
\ee
It is necessary to introduce the covariant coordinates through the shift 
(\ref{shift}), because, unlike for a commutative gauge theory, 
on a noncommutative space 
$\delta (x^i \star \Psi)=i x^i \star \Lambda \star \Psi \neq i
\Lambda \star x^i \star \Psi$.

The gauge parameter $\Lambda$ is required to transform under $\d$ as
\be
\delta \Lambda=i \Lambda \star \Lambda~. \label{swl}
\ee

The SW map is constructed by considering the noncommutative gauge potential 
$A^i=A^i(a_j, \del^n a_j)$ and the noncommutative gauge parameter
$\Lambda=\Lambda(\lambda,\del^n \lambda,a_i,\del^n a_i)$ as
functions of the commutative gauge potential $a_i$, the commutative gauge
parameter $\lambda$ and their derivatives. 
The functional dependence is defined by the equations (\ref{swa}) and
(\ref{swl}).
Notice that throughout this section the convention is used, that 
quantities with capital letters such as $A^i$, $\Lambda$ refer to the 
noncommutative theory, while quantities such as $a_i$, $\lambda$ with lower
case letters refer to the corresponding commutative theory.

In order to solve the equations (\ref{swa}) and (\ref{swl}), a cohomological
method can be used, as it has been discussed in 
\cite{BCPVZ}. Even if $\theta^{ij}$ in (\ref{deft}) is linear and 
not constant, in this case this technique still works. Here, the main
results of \cite{BCPVZ} are briefly recalled.

The gauge parameter $\Lambda$ is promoted to a ghost
field and $\delta$ to a BRST operator, which satisfies
\be
\d^2=0, \qquad [\d,\del_i]=0, \qquad
\d (f_1 f_2)=(\d f_1) f_2 +(-1)^{deg(f_1)} f_1 (\d f_2),
\label{prop} \ee
where $deg(f)$ gives the ghost number of the expression $f$.

The noncommutative gauge parameter and gauge potential can be expanded in 
powers of $\t^{ij}$:
\ba
\Lambda=\lambda+\Lambda^{(1)}+\ldots, \qquad A^i=\t^{ij} a_j +{A^i}^{(2)}+\ldots
\ea
In this formalism the index of the lowest order term of the gauge potential
is raised with $\t^{ij}$, so that the first non-trivial order in the expansion
of $A^i$ is the second.

The equations (\ref{swl}) for $\Lambda$ and (\ref{swa}) for $A^i$ become
\be
\begin{array}{ll}
\delta \Lambda^{(n)} - i \{ \lambda, \Lambda^{(n)} \} = M^{(n)} ~,\\ 
\delta {A^i}^{(n)} - i [ \lambda, {A^i}^{(n)} ] = {U^i}^{(n)} ~,
\end{array}
\label{lan}
\ee
where $M^{(n)}$ and ${U^i}^{(n)}$ collect all the terms of order $n$ which
do not contain $\Lambda^{(n)}$ and ${A^i}^{(n)}$ respectively.
In order to solve (\ref{lan}) it is useful to introduce the new operator
$\Delta$
\be
\Delta=\left \{\begin{array}{ll}
\d -i \{\lambda, \cdot\} & \textrm{on odd quantities,} \\
\d -i [\lambda,\cdot ] &  \textrm{on even quantities,}\\
\end{array}
\right.
\ee
which is nilpotent, obeys the same Super-Leibniz rule as $\delta$, and
commutes with the covariant derivative 
\be
D_i=\left\{ \begin{array}{ll}
\del_i \cdot-i\{a_i, \cdot\} &\textrm{on odd quantities,} \\
\del_i \cdot-i[a_i, \cdot]&  \textrm{on even quantities.}
\end{array}
\right.
\ee
With the notation $b_i \equiv \del_i \lambda$, it can be seen that 
$\Delta a_i=b_i$, $\Delta b_i=0$.
It is not possible to invert the nilpotent operator $\D$ to solve
(\ref{lan}), but, following \cite{BCPVZ}, if an homotopy operator $K$ 
is introduced such that
\be
K \D+\D K=1,
\ee
then for a quantity $m$ such that $\Delta m=0$
the equation $\Delta f=m$ is solved by $f=K \: m+s$ for any $s$ such that
$\Delta s=0$.

As in the case of constant $\theta^{ij}$ to construct $K$, the first step is
to introduce the operator $L$, which obeys the Super-Leibniz rule and satisfies
\be
L \: b_i=a_i, \qquad L \: a_i=0,
\ee
then define $K=D^{-1} L$, where $D^{-1}$ is a linear operator which when
acting on a monomial of total order $d$ in $a$ and $b$ multiplies that 
monomial by $1/d$. Both $L$ and $\delta$ do not act on $\t^{ij}$, i.e.
$\delta  \t^{ij}=0$ and $L \t^{ij}=0$.

The nilpotency of $\Delta$ implies the consistency condition for
$M^{(n)}$ and $U^{(n)}$
\be
\Delta M^{(n)}=0, \qquad \Delta {U^i}^{(n)}=0.
\label{cons}
\ee

There are no corrections to the first order term $\Lambda^{(1)}$ of 
$\Lambda$ due to the time-dependence of $\t^{ij}$, because the Kontsevich star
product (\ref{kont}) does not contain terms in $\del_i \t^{kl}$.
Therefore, to the first order the known expression
$\Lambda^{(1)}=\frac{1}{4} \t^{kl} \{b_k,a_l\}$ \cite{SW} is recovered.

However, there is a correction to the second order term. If $\Lambda^{(2)}$
is split in $\Lambda^{(2)}={\Lambda'}^{(2)}+{\Lambda''}^{(2)}$ with
$\Lambda'^{(2)}$ denoting the known terms (see e.g. \cite{GoHa}, 
\cite{JMSSW}, \cite{BCPVZ}) which do not depend on derivatives
$\del_k \t^{ij}$ of $\t^{ij}$, and ${\Lambda''}^{(2)}$ denoting the
correction due to the fact that $\t^{ij}$ is not constant, then
\ba
{\Lambda''}^{(2)}=-\frac{1}{4} \t^{ij} \del_j \t^{kl} \left( \frac{1}{6} (
\left\{a_i,\{b_k,a_l\}\right\}+i[D_i a_k,b_l]-i[D_i b_k,a_l])\right. 
\label{corrl} \\
+\left. \frac{1}{9} \left([[a_i,b_k],a_l]-[[a_i,a_k],b_l]\right)\right)~, \nn
\ea
which in this case becomes
\ba
{\Lambda''}^{(2)}&=&\frac{1}{4}\widetilde R^2 x^+ \left( \frac{1}{6} (
\left\{a_z,\{b_{x^-},a_z\}\right\}+i[D_z a_{x^-},b_z]-i[D_z b_{x^-},a_z])
\right. \\
&&+\frac{1}{9} \left([[a_z,b_{x^-}],a_z]-[[a_z,a_{x^-}],b_z]\right)
-\frac{1}{6} (\left\{a_z,\{b_z,a_{x^-}\}\right\} \no
&&\left. +i[D_z a_z,b_{x^-}]-i[D_z b_z,a_{x^-}])
- \frac{1}{9} \left([[a_z,b_z],a_{x^-}]-[[a_z,a_z],b_{x^-}]\right)\right).
\nn
\ea
The expression (\ref{corrl}) for ${\Lambda''}^{(2)}$ is determined by solving
the equation
\be
\Delta {\Lambda''}^{(2)}={M''}^{(2)} \equiv
-\frac{1}{4} \t^{ij} \del_j \t^{kl} 
\left(\ha [b_i,\{b_k,a_l\}]+\frac{1}{3}\{iD_i b_k-[a_i,b_k],b_l\}\right)~,
\label{orswl}
\ee
since $M^{(n)}$ can be also split in a part ${M'}^{(n)}$ which does not 
depend on derivatives of the noncommutativity parameter and a correction
${M''}^{(n)}$ due to the fact that $\t^{ij}$ is not constant and then the
two equations $\Delta  {\Lambda'}^{(n)}={M'}^{(n)}$ and 
$\Delta  {\Lambda''}^{(n)}={M''}^{(n)}$ can be solved separately.
In particular ${M'}^{(n)}$ satisfies the consistency condition (\ref{cons}) 
$\Delta {M'}^{(2)}=0$ by itself, therefore ${M''}^{(2)}$
has to satisfy it by itself as well.
This is ensured by the Jacobi identity (\ref{jacobi}) for $\t^{ij}$.
\be
\Delta {M''}^{(2)}=\frac{1}{6} \t^{ij} \del_j \t^{kl}(b_i b_k b_l
+ b_k b_l b_i+b_l b_i b_k)=0.
\ee

An analogous computation can be done for the gauge potential.
Splitting again ${A^i}^{(n)}={{A'}^i}^{(n)}+{{A''}^i}^{(n)}$
in a part ${{A'}^i}^{(n)}$ which does not depend on $\del \t$ and in a part
${{A''}^i}^{(n)}$ which does, then the result ${{A'}^i}^{(2)}=
-\frac{1}{4} \t^{ij} \t^{kl} \{a_k,\del_l a_j+F_{lj}\}$ \cite{SW} is
recovered, with $F_{lj}$ the field strength. The lowest order 
correction due to the fact that $\t^{ij}$ is time-dependent is found to be
\be
{{A''}^i}^{(2)}=-\frac{1}{4} \t^{kl} \del_l \t^{ij} \{a_k,a_j\},
\label{a1}
\ee
which in this case means
\be
{{A''}^i}^{(2)}=
\left\{
\begin{array}{ll}
{1 \over 4} \widetilde R^2 x^+ \{a_z,a_z\} & \mbox{for } i=x^-, \\[1em]
-\frac{1}{4} \widetilde R^2 x^+ \{a_z,a_{x^-}\} & \mbox{for } i=z, \\[1em]
0 & \mbox{otherwise.}
\end{array}
\right.
\label{corra1}
\ee
The result (\ref{a1}) can be obtained by applying the homotopy operator 
$K$ to ${{U''}^i}^{(2)}$ and thus solving the equation
\be
\Delta {{A''}^i}^{(2)}={{U''}^i}^{(2)} \equiv
\frac{1}{4} \t^{ij} \del_j \t^{kl} 
\{b_k,a_l\}-\ha \t^{kl}  \del_l \t^{ij}\{b_k,a_j\}.
\ee
Again, the consistency condition $\Delta {{U''}^i}^{(2)}=0$
is guaranteed by the Jacobi identity for $\t^{ij}$.

Notice that the expressions (\ref{corrl}) and (\ref{a1}) for ${\Lambda''}^{(2)}$
and ${{A''}^i}^{(2)}$ are valid for a general non-abelian gauge group,
but they reduce to the known expressions given in \cite{JSW} in the case
of an abelian gauge theory.

For simplicity the correction to the next order of $A^i$ due to the fact
that $\t^{ij}$ is not constant is computed here only in the
abelian case, even though in principle it would be possible to solve
it even in the more general non-abelian case. The result is
\ba
{{A''}^i}^{(3)}&=&\frac{1}{4}\left( -\frac{4}{3} \t^{ij} \del_j \t^{rs}
\t^{kl} a_k a_r (\del_l a_s)_S \right. +\t^{sj} \del_j \t^{ir} \t^{kl} a_k a_r 
\left(f_{ls}+\frac{2}{3} (\del_l a_s)_S\right) \no
&&\left. + \t^{rj} \del_j \t^{si} \t^{kl} a_r a_k \left(2 f_{sl} -\frac{4}{3} 
(\del_l a_s)_S \right)\right)  \label{a2} \\
&& +\frac{1}{12} \t^{ij}  \t^{kl} \del_l \t^{rs} a_k a_s \left(5 f_{jr}
-2 (\del_j a_r)_S\right)+\frac{1}{6} \t^{kl}  \del_l \t^{rs} \del_s \t^{ij}
a_k a_r a_j~, \nn
\ea
where $(\del_l a_s)_S=\frac{1}{2} (\del_l a_s +\del_s a_l)$ is the symmetrized
derivative of the gauge potential and $f_{ls}=\del_l a_s-\del_s a_l$ is the 
abelian field strength. 
The expression (\ref{a2}) can be found by applying K to
\be
{U^i}^{(3)}=\t^{ij} \del_j \Lambda^{(2)} -
\ha \t^{kl} \left\{ b_k,\del_l {A^i}^{(1)} \right\}-\ha \t^{kl} \left\{ \del_k 
\Lambda^{(1)},\del_l (\t^{ij} a_j )\right\}~.
\ee
Again, the consistency condition $\Delta {{U''}^i}^{(3)}=0$ is verified because
of the Jacobi identity for $\t^{ij}$. Moreover, it is necessary to apply the
constraints $\del_i b_j=\del_j b_i$ and $\del_i a_j=\ha f_{ij}+(\del_i a_j)_S$
by hand, as explained in \cite{BCPVZ}, in order to obtain (\ref{a2}).

The results (\ref{corrl}), (\ref{a1}) and (\ref{a2}) 
for ${\Lambda''}^{(2)}$, ${{A''}^i}^{(2)}$ and ${{A''}^i}^{(3)}$ respectively
are valid in the general case of a linear $\theta^{ij}$ which satisfies the
Jacobi identity.

\section{Conclusions}

Time-dependent backgrounds have recently attracted much attention
in \linebreak string theory \cite{FS}-\cite{BaHu}. 
Although it can be difficult to
interpret singular time-dependent backgrounds in string perturbation theory
\cite{HoPo, DoHu}, here only the scaling limit at the level of the
corresponding noncommutative gauge theory is considered.

The results for the SW map in section 4 are a generalization to higher 
orders in $\t^{ij}$ of the formula (105) in \cite{DRRS}.
For an algebra related to (\ref{defx}), (\ref{deft}), the corresponding
noncommutative gauge theory and its relations to matrix theory are studied 
in \cite{LNR}.

The equivalence map in section 3 could be used in principle, where it is not
singular, i.e. outside $x^+=y^+=0$, to map the results known for the
case of constant $\t^{ij}$ to the case of the algebra (\ref{deft}),
(\ref{defx}).

\section*{Acknowledgments} 

BLC would like to thank P.~Aschieri, L.~Dolan, O.~Ganor, B.~Jur\v{c}o, 
L.~M\"oller, J.~Wess, B.~Zumino for very useful discussions.
BLC was supported in part by US Department of Energy, grant 
DE-FG 05-85ER-40219/Task A. BLC would like to thank B.~Zumino for the
invitation to Berkeley and J.~Wess for the invitation to Munich.

\end{document}